\documentclass[fleqn,12pt,twoside]{article}
\usepackage[headings]{espcrc1}

\newcommand{\vecr}{\mbox{\boldmath $r$}}

\newcommand{\veck}{\mbox{\boldmath $k$}}

\def\m@thcombine#1#2{%
  \setbox0=\hbox{$#1$}
  \setbox1=\hbox{$#2$}
  \ifdim\wd0>\wd1
    \setbox0=\hbox to\wd1{\hss\box0\hss}
  \else
    \setbox1=\hbox to\wd0{\hss\box1\hss}
  \fi
  \mathop{\vcenter{
    \offinterlineskip\box0\box1}}}
\def\lesim{\m@thcombine<\sim}
\def\gesim{\m@thcombine>\sim}

\readRCS
$Id: espcrc1.tex,v 1.2 2004/02/24 11:22:11 spepping Exp $
\usepackage{graphicx}

\newcommand{\AmS}{{\protect\the\textfont2
  A\kern-.1667em\lower.5ex\hbox{M}\kern-.125emS}}

\title{Pairing collectivity in medium-mass neutron-rich nuclei
near drip-line
        \thanks{
This work was supported by the Grant-in-Aid for Scientific
Research (No. 17540244) from the Japan Society for the Promotion
of Science.}
}

\author{M. Matsuo\address[N]{Department of Physics, Faculty of
Science Niigata University, 
Niigata 950-2181, Japan},
Y. Serizawa\addressmark,
 and K. Mizuyama\addressmark
}
       
\runtitle{Pairing collectivity in medium-mass neutron-rich nuclei}
\runauthor{M. Matsuo}

\begin{document}

\maketitle

\begin{abstract}
We look for collective excitations originating from the 
strong surface pairing in unstable nuclei near the neutron drip-line. 
The soft dipole excitation is such a pairing mode as 
it exhibits a character of core-vs-dineutron motion.
Possibility of the hydrodynamic phonon mode (the Anderson-Bogoliubov 
mode) is also discussed.
\end{abstract}

\section{Introduction}

The pair correlation is known to play central roles in
characterizing various structure aspects of 
nuclei\cite{Dean03}. As far as nuclei near the stability line are
concerned, the energy scale of the correlation, the pair
gap $\Delta \sim 1-2$ MeV, is 
much smaller than the other fundamental energy scales,
the Fermi energy $e_F \sim 40$ MeV, the shell gap $\sim$ several MeV 
and the nucleon separation energy $\sim$ 8 MeV. This situation
can be regarded as that of the weak coupling pairing to which
the conventional BCS models can be successfully applied.
The correlated nucleon pair in this case is reasonably described as
a pair coupled to the total angular momentum $I=0$ consisting
of a small number of single-particle $j$-shell orbits around the Fermi energy.

Unstable nuclei near the neutron drip-line are in a very
different situation. The neutron separation energy 
becomes comparable with or even smaller than the typical pair gap. 
The single-particle shell 
tends to lose its sense since the continuum 
orbits take part in.  Moreover neutrons in the low density external region
associated with skin and/or halo feel stronger n-n attraction due to the
momentum dependence of the nuclear force.
One can expect then that
the pair correlation in this 'extreme' condition
may differ from that in stable nuclei. 
An important clue is
the di-neutron correlation (the spatially compact correlated neutron
pair), which has been investigated intensively 
for light two-neutron halo nuclei  
$^{11}$Li and
$^6$He\cite{Nakamura,Hansen,Bertsch91,Ikeda,Zhukov,Barranco01,Aoyama,Hagino05}.
In addition our previous study suggests that 
the di-neutron correlation persists also in medium-mass neutron-rich
nuclei containing more than two weakly bound neutrons\cite{Matsuo05}. 

In this presentation, we first demonstrate, 
by using a calculation
for uniform neutron matter\cite{Matsuo06}, that 
the di-neutron correlation is naturally expected in
low density region. 
We shall interprete 
the di-neutron correlation in terms of
the BCS-BEC crossover\cite{Leggett,Nozieres,Randeria,Engelbrecht}
which takes place in between the weak coupling pairing
of the conventional BCS theory and the strong coupling pairing
leading to a Bose-Einstein condensate of correlated pairs.
In other words, the di-neutron correlation is an
signature of the strong coupling pairing.
If the strong coupling pairing is the case, 
it will influence excitation properties of nuclei. 
This is the issue which we discuss in the latter half of the presentation.
We look for another manifestation of the strong coupling pairing 
by analyzing dipole and quadrupole responses of medium-mass 
neutron-rich nuclei.

\section{Low density uniform matter and the BCS-BEC crossover}

Let us discuss properties of the neutron pair correlation in 
uniform neutron matter at low density\cite{Matsuo06}. We employ
results of the BCS approximation applied to the bare nuclear force.
The G3RS force is adopted, and the effective mass consistent with 
the Gogny Hartree-Fock is used. The calculation itself is essentially
the same as other BCS calculations using the bare forces
\cite{Dean03}.
  
Within this scheme the gap and the number equations can be
solved exactly by a numerical method as a function of neutron
density $\rho=k_F^3/3\pi^2$, where $k_F$ is the
Fermi momentum. We can 
analyze spatial structure of the pair correlation by means of
the wave function of the correlated neutron pair
(the Cooper pair), which is given by
\begin{equation}\label{Cooper-eq}
\Psi_{pair}(r) \propto
  \left<\Phi_0\right|\psi^\dagger(\vecr'+\vecr/2\uparrow)
\psi^\dagger(\vecr'-\vecr/2\downarrow)\left|\Phi_0  \right>
= 
{1 \over (2\pi)^3}
\int d \veck  u_kv_k e^{i\veck\cdot\vecr}
\end{equation}
where $u_k,v_k$ are the u,v-factors in the momentum representations.
As a measure of the spatial structure, 
the r.m.s. radius of the Cooper pair or the coherence length 
\begin{equation}\label{rms-eq}
\xi= \sqrt{\left<r^2\right>}, \hskip 10mm
\left<r^2\right>= \int d\vecr r^2 |\Psi_{pair}(r)|^2
\end{equation}
is evaluated from the Cooper pair wave function.

Fig.1 shows examples of the neutron Cooper pair wave function (its
square modulus $r^2|\Psi(r)|^2$) in neutron matter 
at the density $\rho/\rho_0=$1, 1/2, 1/8 and 1/64
($\rho_0$ being the neutron density in the
normal nuclear matter given by 
$\rho_0=k_{F,0}^3/3\pi^2$ with $k_{F,0}=1.36$ 
fm$^{-1}$ ). 
The pair gap $\Delta$ evaluated at the Fermi energy 
and the coherence length $\xi$ are also shown in the figure.
An important observation is that the size of
the neutron Cooper pair is as small as $\xi \sim 5$ fm
at the low density $\rho/\rho_0=1/8 - 1/64$. This is in
contrast to the cases around the normal density 
$\rho/\rho_0=1 - 1/2$ where the coherence
length becomes considerably large $\xi \gesim 10^{1}$fm.
It is seen in Fig.1 that the profile of the Cooper pair
wave function also varies significantly with the density.
At the density $\rho/\rho_0=1/8 - 1/64$, the 
profile resembles to that of a bound state wave function in free
space while the wave function at the normal density deviates
significantly from that of the free-space bound state.

\begin{figure}[tb]
\includegraphics[width=5.5cm,angle=-90]{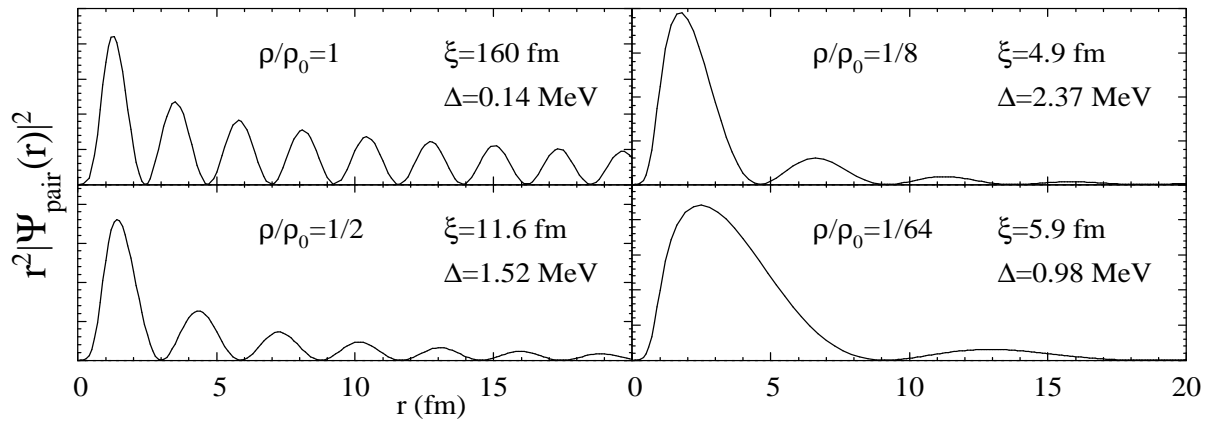}
\caption{The probability distribution $r^2|\Psi_{pair}(r)|^2$
of the wave function of a neutron Cooper pair in low
density neutron matter.
The horizontal axis is the relative coordinate $r$ between
the partner neutrons. 
}
\label{fig:fig1}
\end{figure}

Implications of the spatial di-neutron correlation
at the low density can be clarified by considering a possible link to
the BCS-BEC crossover\cite{Leggett,Nozieres,Randeria,Engelbrecht}. 
The crossover is
a general phenomenon which occurs in superfluid/superconducting 
Fermion systems as the attractive interaction 
among the partner Fermions is varied from the weak to strong
cases. The crossover has been observed recently in an ultra cold
Fermion gas in a trap\cite{Regal}.
The weak coupling case corresponds to the conventional
electron superconductivity in metals, and it is often called the
weak coupling BCS since it is the situation assumed in
the original BCS theories. In this case, 
the pair gap $\Delta$ is much smaller than the Fermi energy
$e_F$, and  the coherence
length $\xi$ (the size of the Cooper pair) is much larger than
the average inter-particle distance $d=\rho^{-1/3} =3.09k_F^{-1}$.
If the attraction among the pair partners
becomes stronger (the pair gap gets larger accordingly), 
the ratio $\Delta/e_F$ between the pair gap and 
the Fermi energy increases monotonically, and the ratio
$\xi/d$ between the coherence length and the average inter-particle
distance decreases. In the limit of strong coupling, the Cooper pair
coincides with the bound Fermion pair in free space. 
The system is then a Bose-Einstein 
condensate (BEC) of the bound pairs. The transition, though gradual
in nature, from the weak coupling BCS to the strong coupling BEC 
occurs around $\Delta/e_F\sim 1$ and $\xi/d\sim 1$. Reference 
values for the two ratios characterizing the BCS-BEC crossover
are\cite{Engelbrecht}; $\Delta/e_F = 0.21$ and $\xi/d = 1.10$ 
for the boundary to the weak coupling BCS,
$\Delta/e_F = 1.33$ and $\xi/d = 0.19$ 
for the boundary to the strong coupling BEC,
and 
$\Delta/e_F = 0.69$ and $\xi/d = 0.36$ 
for the unitarity limit corresponding to 
the midway of the crossover.

\begin{figure}[tb]
\begin{minipage}[t]{80mm}
\includegraphics[width=5.2cm,angle=-90]{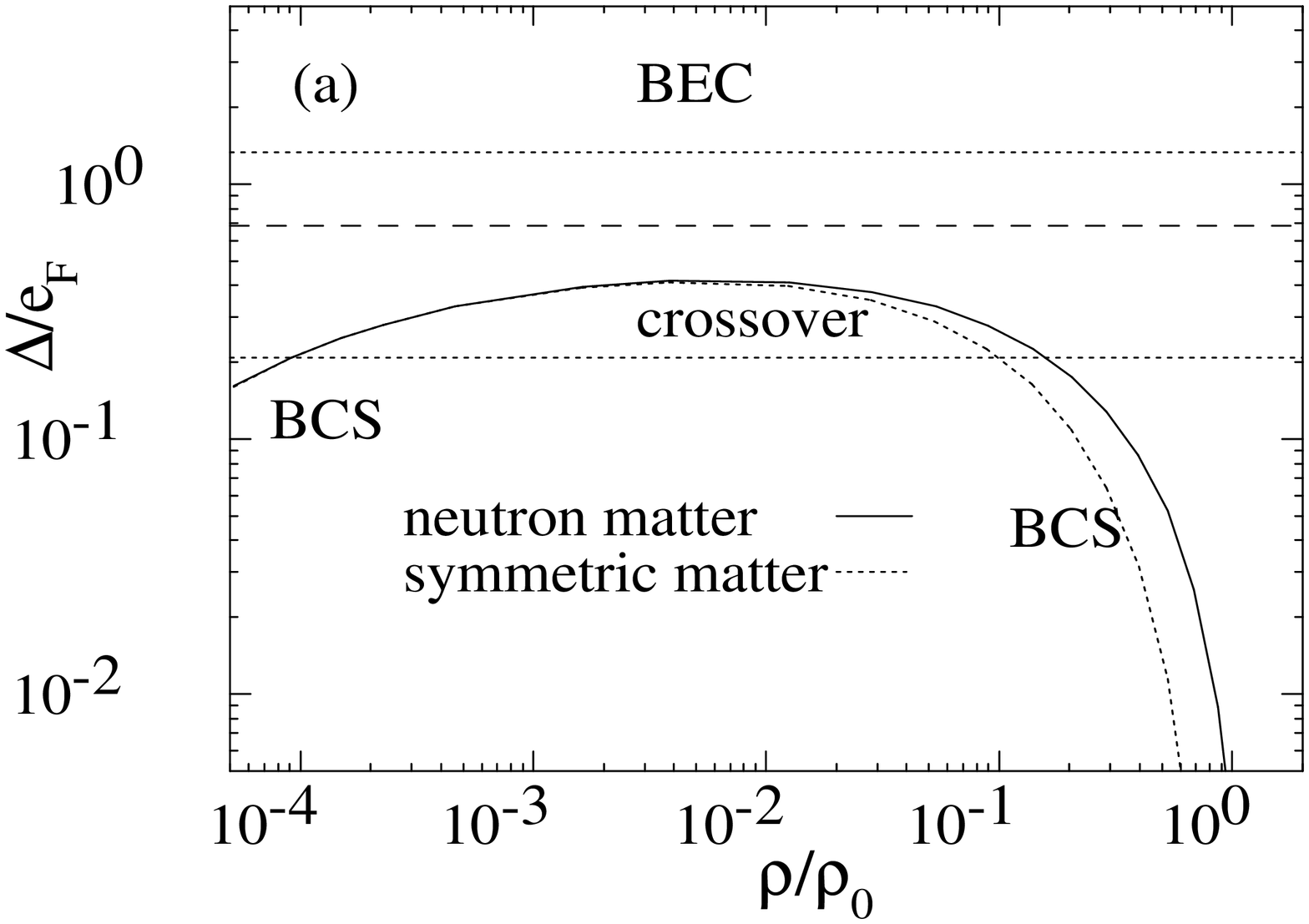}
\end{minipage}
\hspace{\fill}
\begin{minipage}[t]{75mm}
\includegraphics[width=5.2cm,angle=-90]{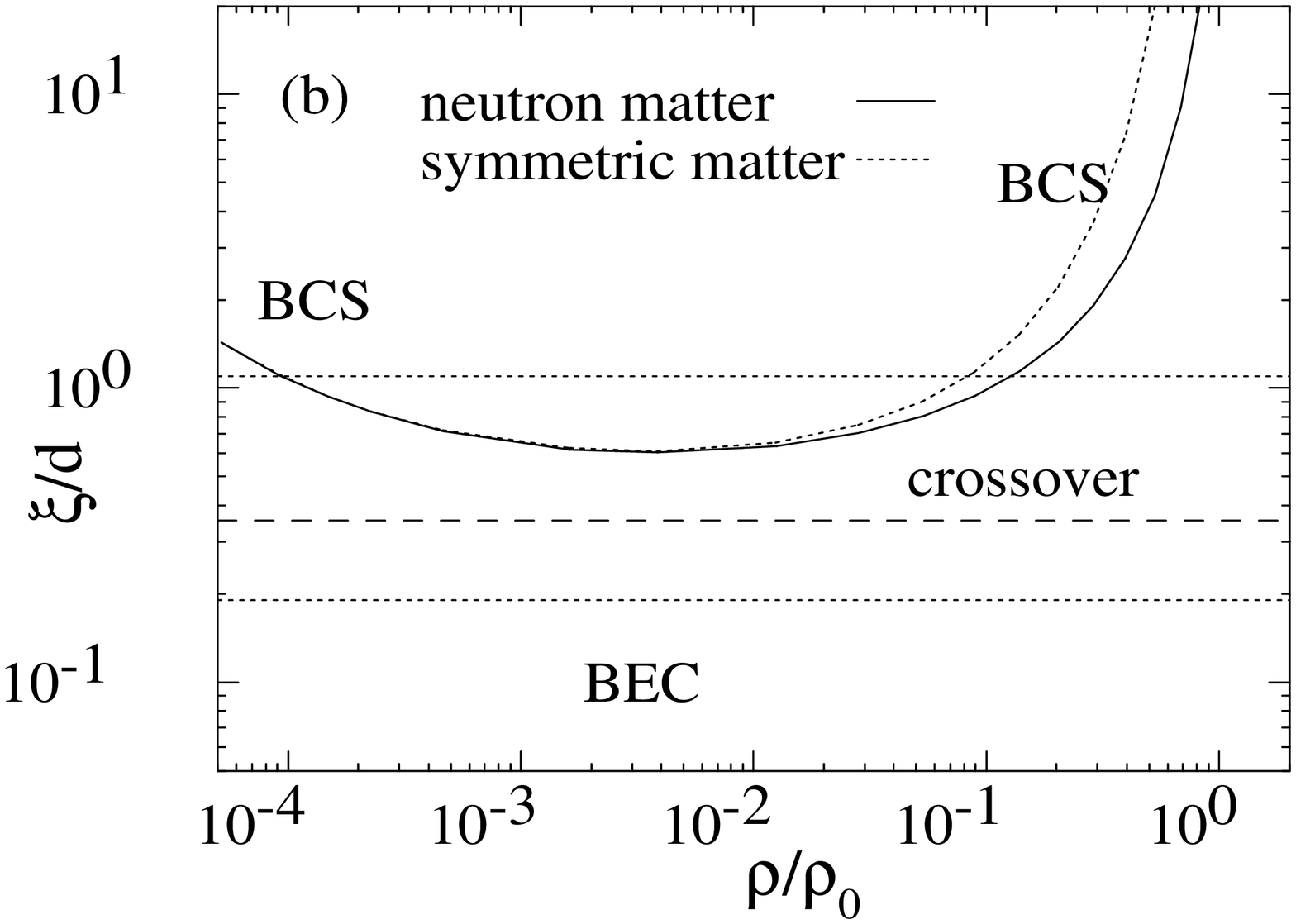}
\end{minipage}
\caption{(a) The ratio $\Delta/e_F$ between the
pair gap $\Delta$ and the Fermi energy $e_F$ as a function
of the neutron density in uniform neutron matter. The boundaries
characterizing the BCS-BEC crossover are drawn by horizontal
dotted and dashed lines. The dotted curve is the ratio 
for uniform symmetric matter. (b) The same as (a) but for the ratio
$\xi/d$ between the coherence length $\xi$ and the
average inter-neutron distance $d$. }
\label{fig:fig2}
\end{figure}

In Fig.2 ploted are the ratios  $\Delta/e_F$ and $\xi/d$ calculated for neutron
matter. Two important features
are seen. Firstly
the ratios strongly vary with the neutron density,
in particular in the range $\rho/\rho_0=1-10^{-1}$.
Secondly the two ratios both indicate that
the system enters the regime of the BCS-BEC crossover at
low density in a wide range $\rho/\rho_0 \sim 10^{-1} - 10^{-4}$.
A strong di-neutron correlation is obvious since 
the size of the neutron Cooper pair $\xi$ is smaller than the 
average inter-neutron distance $d$. 
It should be noted that the strong variation as a function of
the density arises from the fact that the $^{1}S$ interaction 
matrix element or the scattering T-matrix
has strong momentum dependence.

\section{Pairing collectivity in neutron-rich nuclei}

\subsection{Skyrme-HFB plus continuum QRPA approach}

We shall now discuss the pair correlation in neutron-rich nuclei
in the medium-mass region. 
We shall focus on the role of the pair correlation in 
collective responses. For this purpose
we employ the selfconsistent Hartree-Fock-Bogoliubov approach
to describe the ground state and the static mean-fields\cite{DobHFB},
and we also use the continuum QRPA method to describe the
response\cite{Matsuo05,Matsuo01}. 
In the following, we present results for $^{120}$Sn
representative of stable nuclei, and those for $^{158}$Sn, in
which the neutron separation energy is as small as $\sim$1 MeV.

The Skyrme effective interaction with the SLy4 parameter
set is used. 
As the effective
interaction responsible for pairing, we use a density-dependent delta
interaction (DDDI) of the form
$v(\vecr-\vecr')=V_{0,\tau}[\rho(\vecr)]\delta(\vecr-\vecr')$, where
the density dependent interaction strength is given by
\begin{equation}\label{dddi-int-eq}
V_{0,\tau}[\rho(\vecr)]=v_0 \left(1 - \eta \left({\rho_{\tau}(\vecr) \over
				     \rho_c}\right)^\alpha\right), \ \ \
(\tau=n,p), \ \ \ \rho_c=0.08\ {\rm fm}^{-3}.
\end{equation}
Here 
the overall force constant $v_0=-458.4$MeV fm$^{-3}$ is fixed to reproduce the
$^1S$ scattering length $a=-18$fm in free space.
The density dependence factor of the interaction 
enables us to simulate the momentum dependence of
the bare nucleon force. Applying the DDDI to uniform matter, we determine
the remaining parameters as 
$\eta=0.845$ and $\alpha=0.59$, and 
also the cut-off quasiparticle energy ($E_{cut}=60$ MeV) 
so that the DDDI reproduces the pair gap
$\Delta$ and the coherence length $\xi$ obtained 
from the bare force\cite{Matsuo06}.
Direct application of this parameter set to finite nuclei, however,
gives a gap smaller than the experimental values.

\begin{figure}[tb]
\begin{minipage}[t]{70mm}
\includegraphics[width=5.5cm,angle=-90]{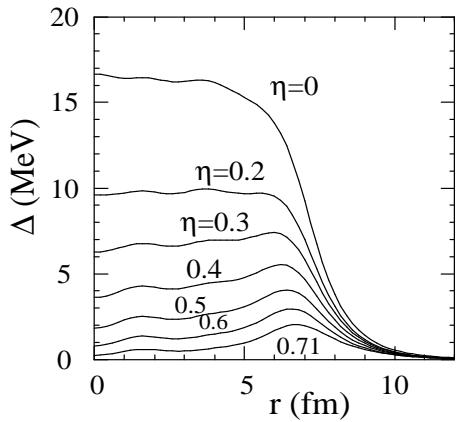}
\end{minipage}
\hspace{\fill}
\begin{minipage}[t]{85mm}
\caption{The neutron pair potential $\Delta(r)$ calculated 
for $^{158}$Sn 
with use of various values of the parameter $\eta$ of the density-dependent
delta interaction.
}
\label{fig:fig3}
\end{minipage}
\end{figure}

If we change the parameter $\eta$, it is possible to control
the effective pairing interaction without changing the scattering
length in free space outside the nuclear surface. 
A smaller value of $\eta$, i.e., weaker density dependence,
gives stronger pairing interaction inside the nucleus. 
We find that $\eta=0.71$ gives an average pair gap
which is comparable to the experimental value
$\Delta \sim 1.1-1.4$ MeV in stable Sn isotopes. If we adopt 
smaller values of $\eta$, the pair gap is further enhanced. 
Figure 3 shows the calculated pair potential 
$\Delta(r)$ for various choices of $\eta$. The case $\eta=0$, implying
a density-independent delta interaction constrained
only by the scattering length, leads to an extremely large
pair potential $\Delta(r) \sim 15$ MeV. This value of the
pair potential is of course too
large compared to the reality. However we exploit this feature
in order to explore what properties could emerge
if the strong coupling pairing dominates in the whole nuclear
volume. 

As the residual interaction to be used in the continuum
QRPA calculation, we use
the same DDDI for the pairing channel while the Landau-Migdal
approximation to the Skyrme interaction is used for the particle-hole
channel. We introduce a renormalization factor to
the particle-hole residual interaction so that 
a spurious center-of-mass mode is placed at the zero-energy.

\subsection{Anderson-Bogoliubov mode associated with strong pairing}

Let us first consider the extreme limit of the strong pairing
($\eta=0$ and $\Delta \sim 15$ MeV) in the stable nucleus
$^{120}$Sn. What kinds of collective mode can be expected in such a case?

\begin{figure}[tb]
\begin{minipage}[t]{80mm}
\includegraphics[width=5.7cm,angle=-90]{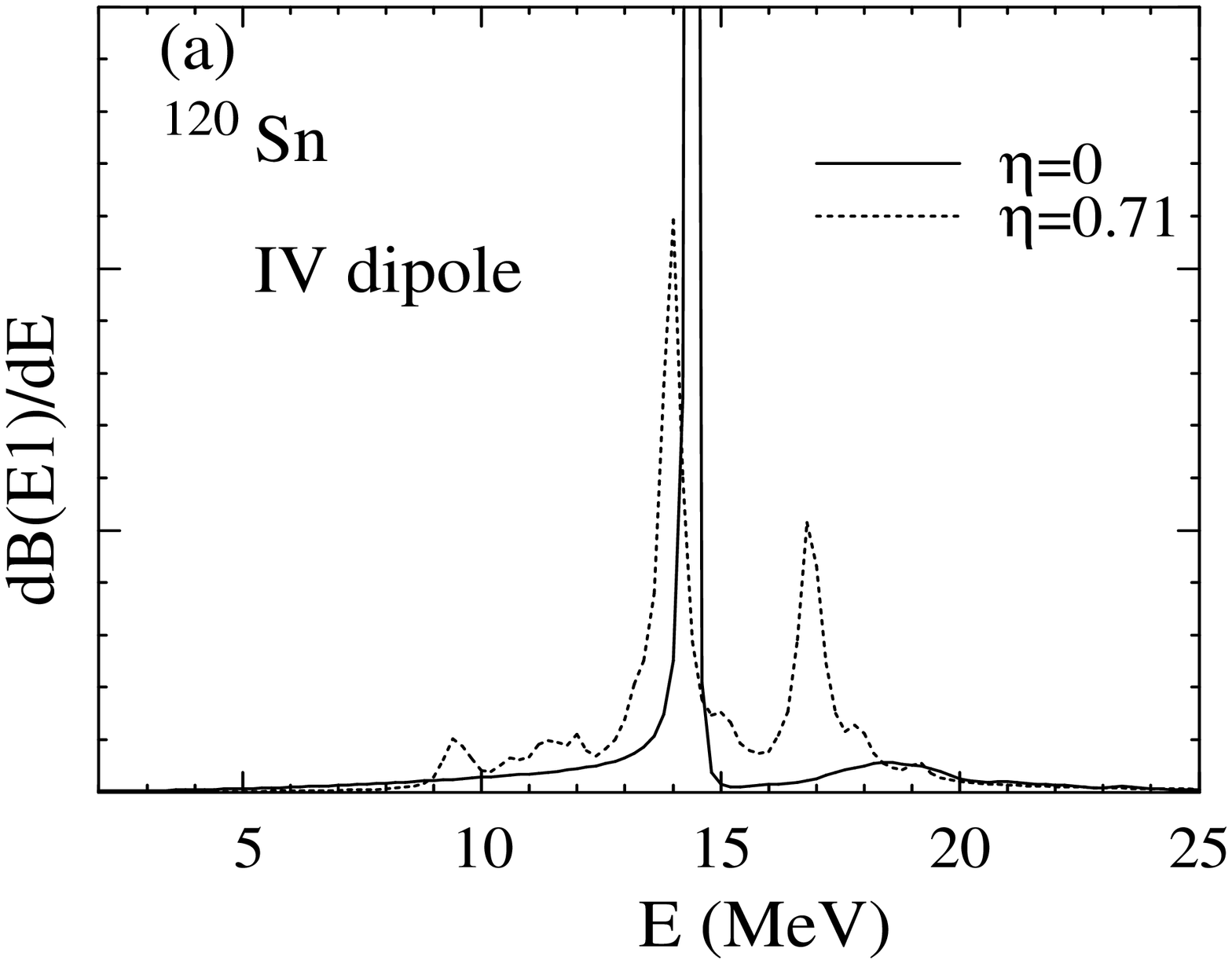}
\end{minipage}
\hspace{\fill}
\begin{minipage}[t]{75mm}
\includegraphics[width=5.5cm,angle=-90]{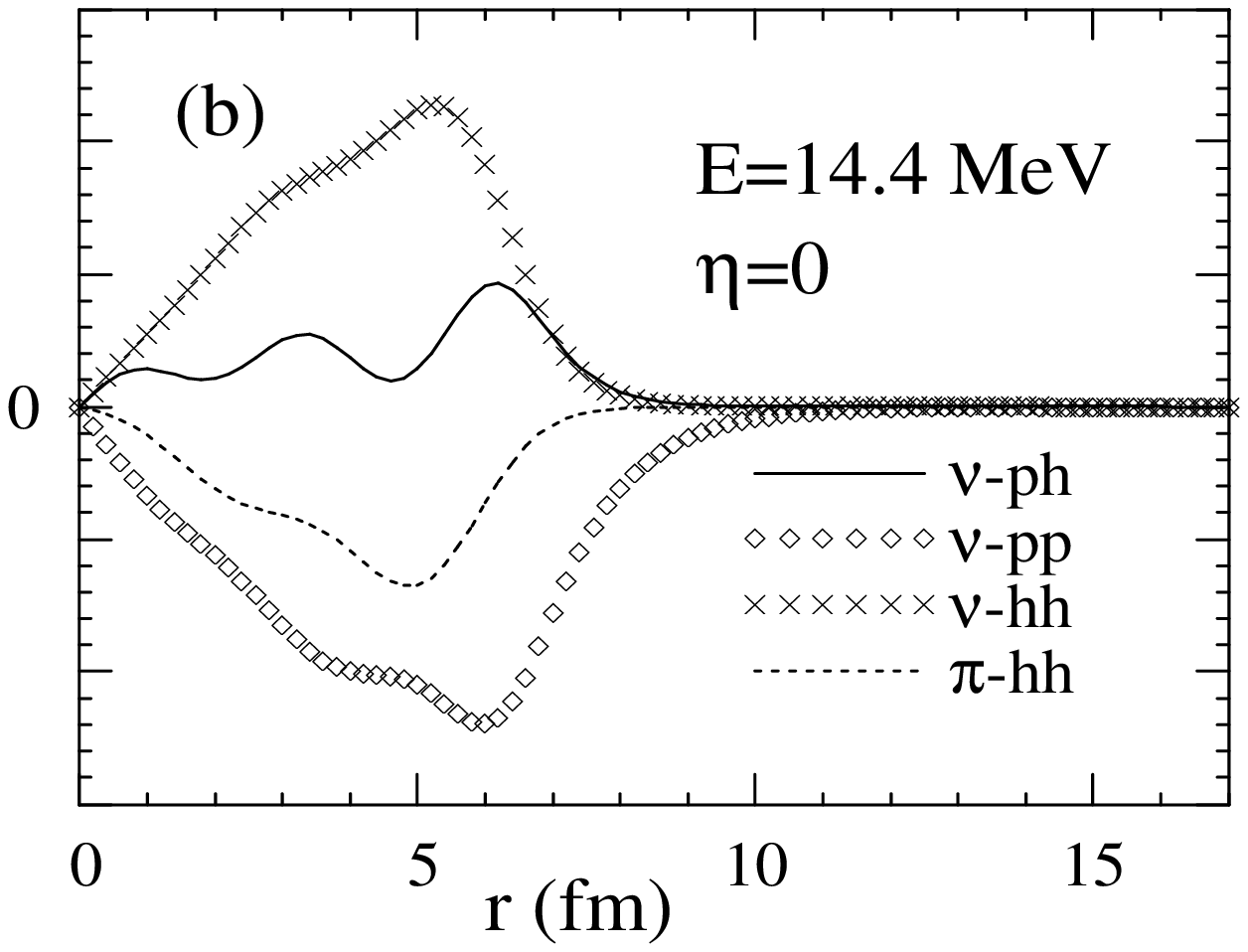}
\end{minipage}
\caption{(a) The E1 strength function 
in $^{120}$Sn calculated using the normal pairing 
($\eta=0.71$, the dotted curve) and the one using the
strong pairing ($\eta=0$, the solid curve).
(b) The transitions densities associated with the 
sharp peak at $E=14.4$ MeV for the strong pairing 
($\eta=0$). The neutron particle-hole transition
density 
$\delta\langle \psi^\dagger(\vecr\sigma)\psi(\vecr\sigma)\rangle$
($\nu-$ph) , 
the neutron particle-pair transition
density 
$\delta\langle \psi^\dagger(\vecr\uparrow)\psi^\dagger(\vecr\downarrow)
\rangle$ ($\nu-$pp), the neutron hole-pair transition
density 
$\delta\langle \psi(\vecr\uparrow)\psi(\vecr\downarrow)\rangle$
($\nu-$hh), 
and the proton particle-hole density
($\pi-$ph) are plotted. 
}
\label{fig:fig4}
\end{figure}

The solid curve in Fig.4(a) shows the E1 strength function
in this case. 
A remarkable feature is that a very sharp resonance 
emerges at $E \sim 14$ MeV. This is quite different
from the widely spread strength distribution, 
the giant dipole resonance, in the case of the normal pairing 
$\eta=0.71$ and $\Delta \sim 1-2$ MeV (the dotted curve in Fig.4(a)).
The transition densities for this sharp resonance (Fig.4(b))
are indeed very different from
those of the isovector giant dipole resonance. 
The largest difference is that the particle-pair transition
density $\delta\langle\psi^\dagger\psi^\dagger\rangle$ and the hole-pair
transition density $\delta\langle\psi\psi\rangle$ are much larger
than the particle-hole transition density
$\delta\langle\psi^\dagger \psi\rangle$. 
This indicates 
that the mode is characterized by an oscillation of the
pair potential $\Delta(\vecr)$, and hence it is a collective mode
associated with the pairing degrees of freedom. The opposite sign between
$\delta\langle\psi^\dagger\psi^\dagger\rangle$ and 
$\delta\langle\psi \psi\rangle$
indicates that the mode accompanies an oscillation in the phase of the
pair potential $\Delta(\vecr)$. 

The collective mode associated with the phase of the pair potential 
$\Delta(\vecr)$ is known 
as the Anderson-Bogoliubov mode
\cite{Anderson,Bogoliubov,Galitskii}.
In uniform superfluid neutral Fermions,  the Anderson-Bogoliubov 
mode is a collective hydrodynamic phonon mode with the dispersion relation
$\omega = cq$, where the sound velocity is given
by $c \approx v_F/\sqrt{3}$, and $q$ and $\omega$ are the momentum and the
frequency of the phonon mode. The phase oscillation 
couples to density oscillation for $q>0$, and hence
it causes hydrodynamic motion of the superfluid.
It is seen in Fig.4(b) that the radial profile of
the calculated transition densities is very much
similar to the Bessel function $j_l(kr)$, a typical
feature of hydrodynamic motion in a spherical system. This indicates
that the sharp resonance is a nuclear Anderson-Bogoliubov mode.
It is also seen in Fig.4(b) that
the transition densities for neutrons and protons have opposite phase.
This mode is therefore
interpreted as an Anderson-Bogoliubov mode 
consisting of two kinds of superfluid oscillating with opposite
phase.
(Because of the presence of two fluids, it differs from 
an Anderson-Bogoliubov mode in a trapped 
superfluid Fermion gas\cite{Baranov,Bruun}.)

It should be remarked here that the features of the Anderson-Bogoliubov mode 
is not seen clearly at weaker pairing. 
In the cases of  $\Delta \sim 3$ MeV ($\eta=0.5$) and 1-2 MeV 
($\eta=0.71$) the transition densities of the high-lying resonance
(not shown here) are characterized not 
by the Anderson-Bogoliubov mode, but rather by the usual giant 
dipole response where the pair transition densities play only minor
roles.

\subsection{Core-vs-dineutron mode in nuclei near the drip-line}

An important question is how the dipole response changes
if we consider neutron-rich nuclei near the drip line. 
Figure 5(a) shows the E1 strength distribution
in $^{158}$Sn for various values of $\eta$, covering from the 
normal pairing case ($\Delta=1-2$MeV, $\eta=0.71$) to the strong pairing case 
($\Delta\sim15$MeV, $\eta=0$). The strength distribution 
in $^{158}$Sn  and those in $^{120}$Sn (Fig.4(a)) are quite different
both in the case of the normal pairing
and in the case of the strong pairing. In the normal pairing
case ($\eta=0.71$), the
strength function exhibits a large low-lying bump 
around $E=2-8$ MeV (the soft dipole excitation) 
beside the high-lying resonance ($E=10-18$ MeV, the giant dipole resonance).
It appears that, with increasing the pair correlation, 
the strength of the low-lying bump develops, and simultaneously
the giant resonance changes its structure.
In the strong pairing case ($\eta=0$), the dipole strength
function exhibits a broad peak around $E=5-15$ MeV.

\begin{figure}[tb]
\begin{minipage}[t]{80mm}
\begin{minipage}[t]{80mm}
\includegraphics[width=5.7cm,angle=-90]{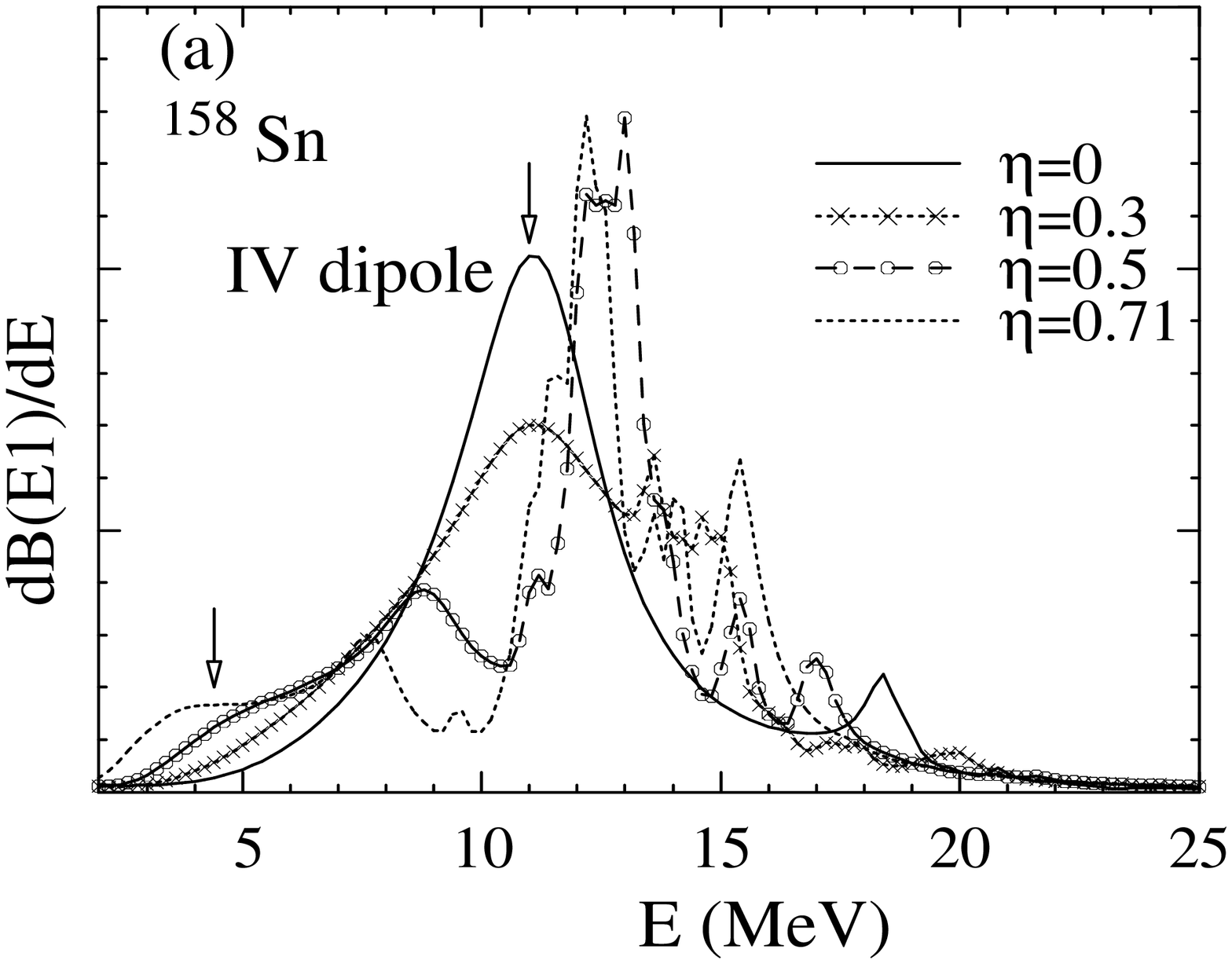}
\end{minipage}
\begin{minipage}[t]{80mm}
\caption{(a) The same as Fig.4, but for $^{158}$Sn. 
The strength distribution obtained with intermediate pairing cases
$\eta=0.3$ and 0.5 are also drawn.
(b) The transition densities at $E=11.0$ MeV
in the strong pairing case ($\eta=0$). (c) 
The transition densities at $E=4.4$ MeV in the normal pairing case
($\eta=0.71$.) }
\label{fig:fig5}
\end{minipage}
\end{minipage}
\hspace{\fill}
\begin{minipage}[t]{75mm}
\begin{minipage}[t]{75mm}
\includegraphics[width=5.5cm,angle=-90]{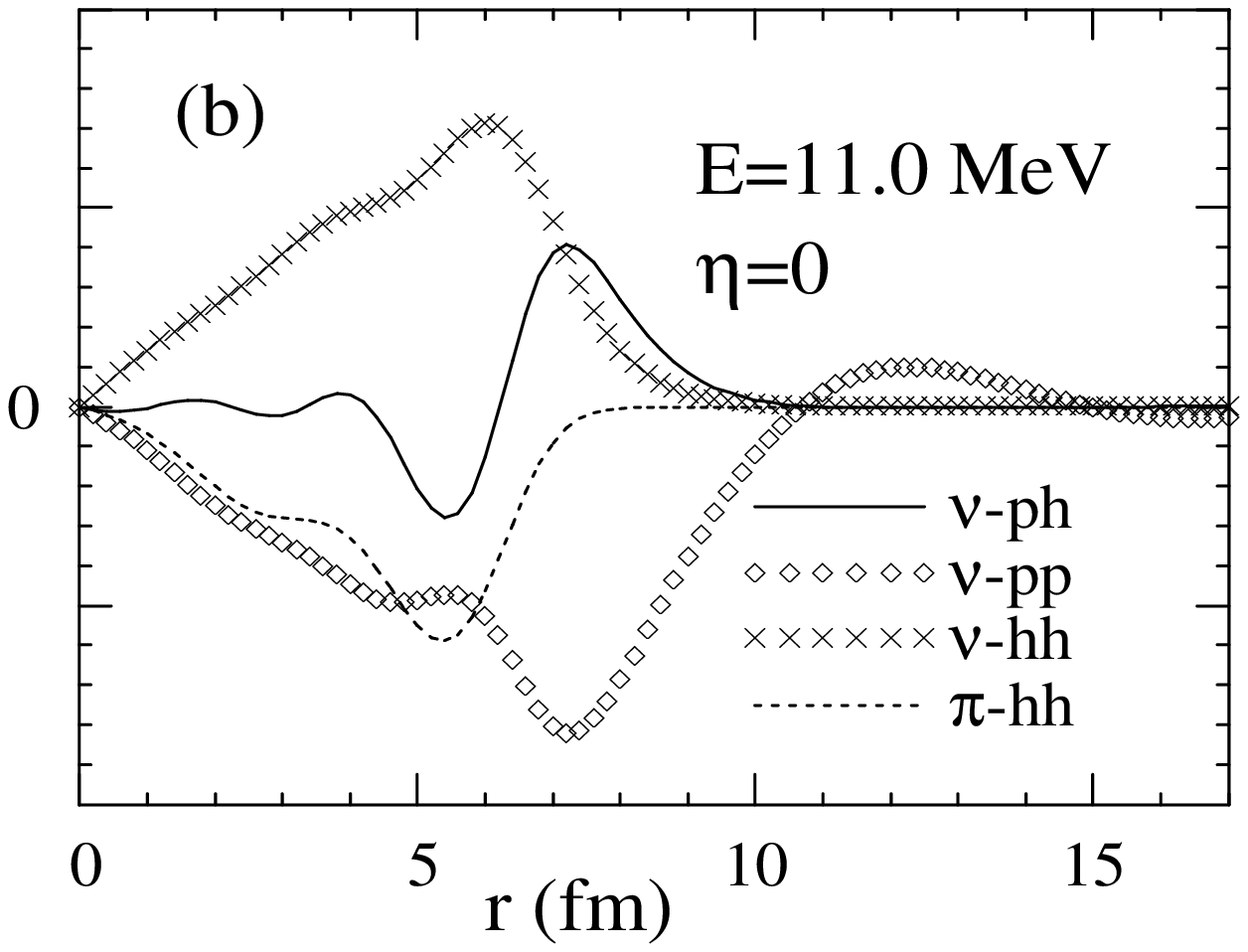}
\end{minipage}
\begin{minipage}[t]{75mm}
\includegraphics[width=5.5cm,angle=-90]{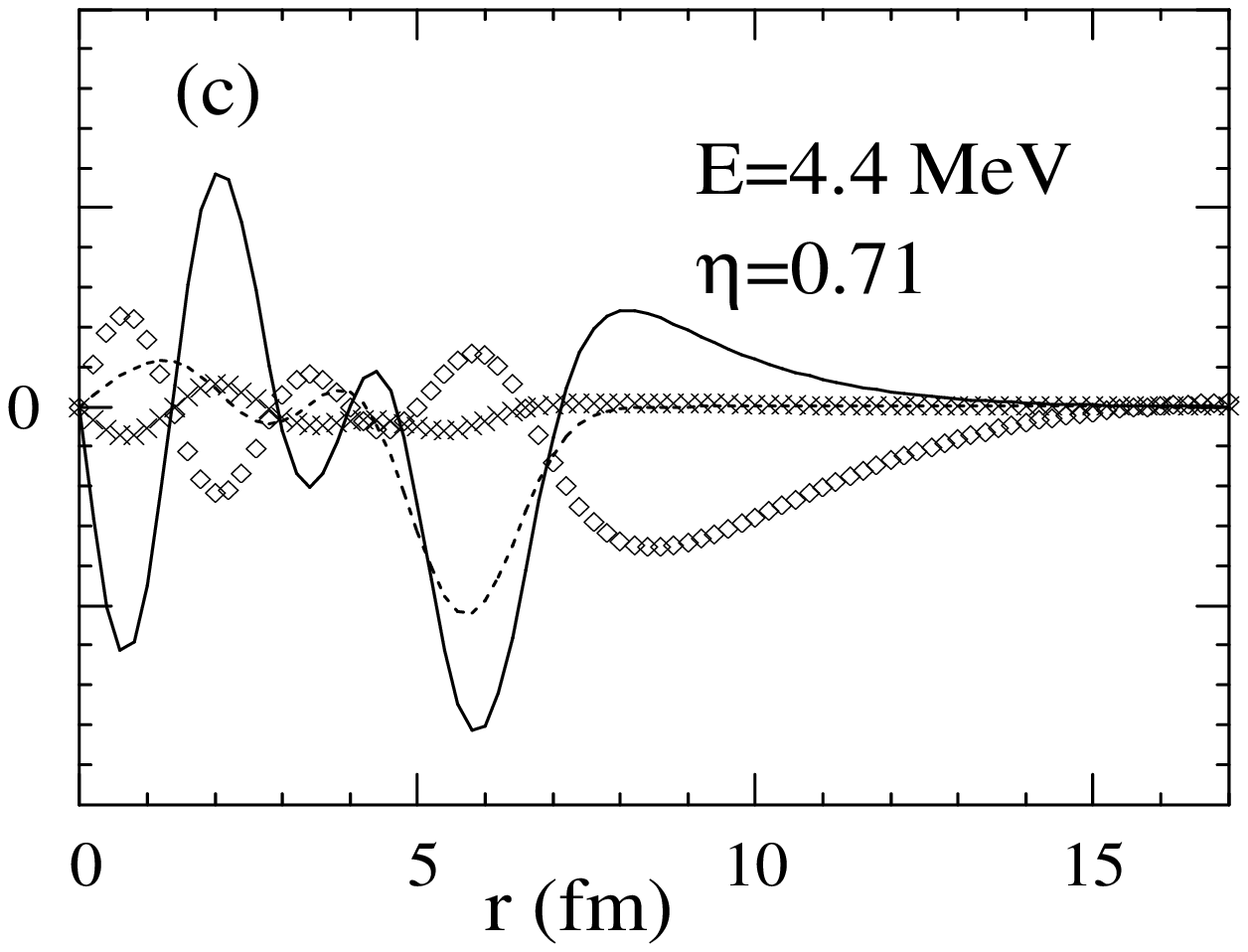}
\end{minipage}
\end{minipage}
\end{figure}

To identify the characters of the dipole modes, we plot in Fig.5(b)  
the transition densities at the broad peak emerging in 
the strong pairing case ($\eta=0, \Delta\sim 15$ MeV).
A clear difference from that of $^{120}$Sn (Fig.4(b)) is 
seen in behaviors around the nuclear surface and in the external
region. In the case of $^{158}$Sn,
the particle-hole transition density 
for neutrons (the solid curve in Fig.5(b)) has
a node at a position near the nuclear surface. Just inside this node, 
neutrons and protons move with the same phase while 
in the external region neutron motion is dominating.
This feature indicates that neutrons in the external region 
move against the dipole motion of the core. It is also noted that the
neutron-pair transition density ($\nu$-pp) has the largest amplitude
in the external region, and it exhibits oscillation
far outside the surface. This can be interpreted as being caused by 
motion of the di-neutrons (a core-vs-dineutron motion), 
which accompanies significant escaping of
di-neutrons from the excitated states\cite{Matsuo05}.

As far as the internal region is concerned, on the
other hand, the transition densities
exhibit the typical features of the Anderson-Bogoliubov mode found
in Fig.4(b). The broad peak around
$E=5-15$ MeV in the strong pairing case has a character 
of a mixture of the Anderson-Bogoliubov mode
and the core-vs-dineutron mode. 

Fig.5(c) is the transition densities for the soft-dipole excitation
seen in the normal pairing case. Focusing on the properties around the
surface and in the external region, we notice that both
Fig.5(b) and (c) exhibit a character of the
core-vs-dineutron motion.
Combining resutls for intermediate pairing strengths
$\eta=0.3$, and 0.5  (not shown here),  we find that 
the core-vs-dineutron mode is always present in the low-lying
part of the dipole strength distribution in the whole
considered range $\eta=0-0.71$ of the parameter $\eta$.
This is quite in contrast to the fact that the Anderson-Bogoliubov mode
emerges only at the very strong pairing $\Delta > 7$ MeV.

\subsection{Quadrupole core-vs-dineutron mode} 

\begin{figure}[tb]
\begin{minipage}[t]{80mm}
\includegraphics[width=5.7cm,angle=-90]{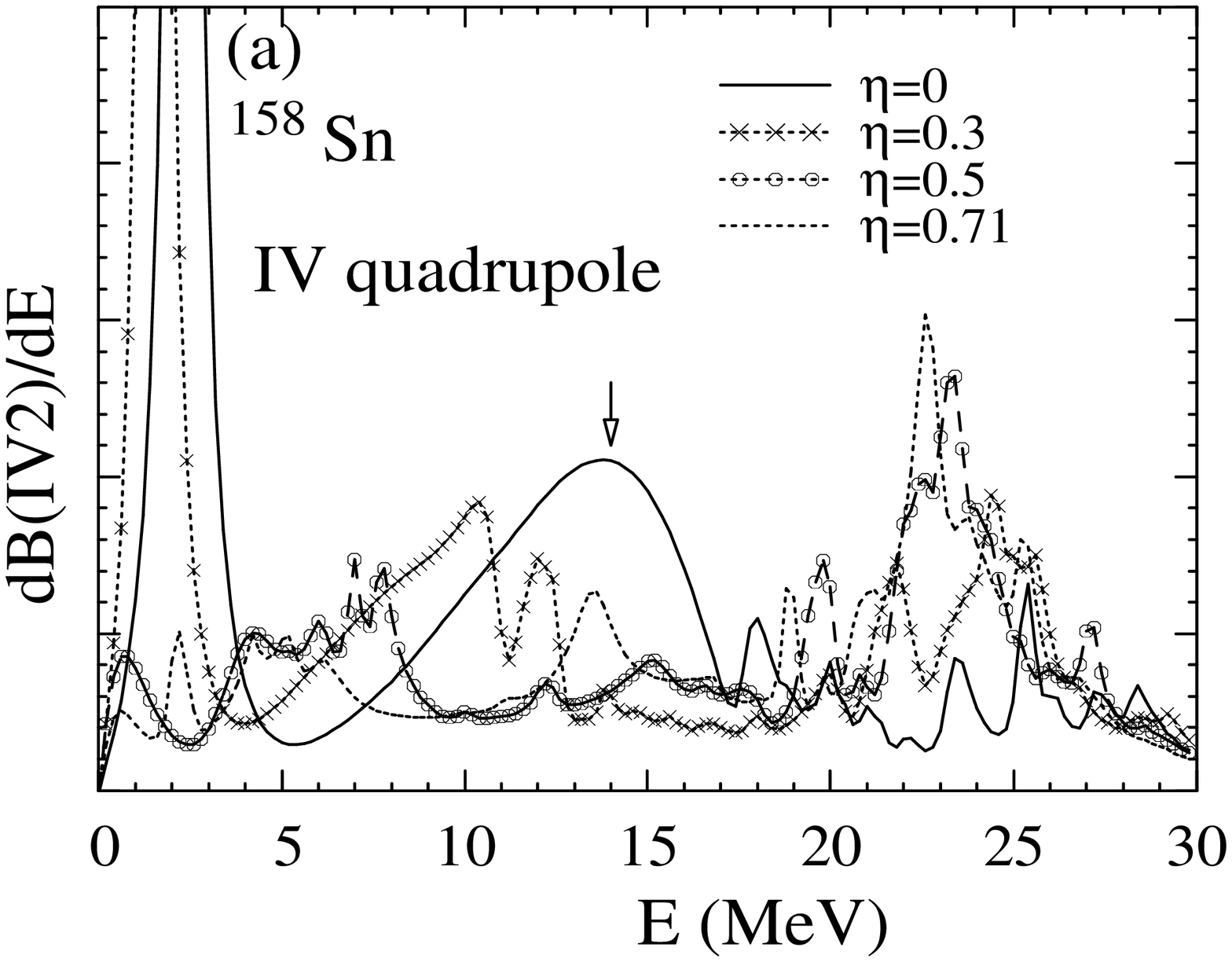}
\end{minipage}
\hspace{\fill}
\begin{minipage}[t]{75mm}
\includegraphics[width=5.5cm,angle=-90]{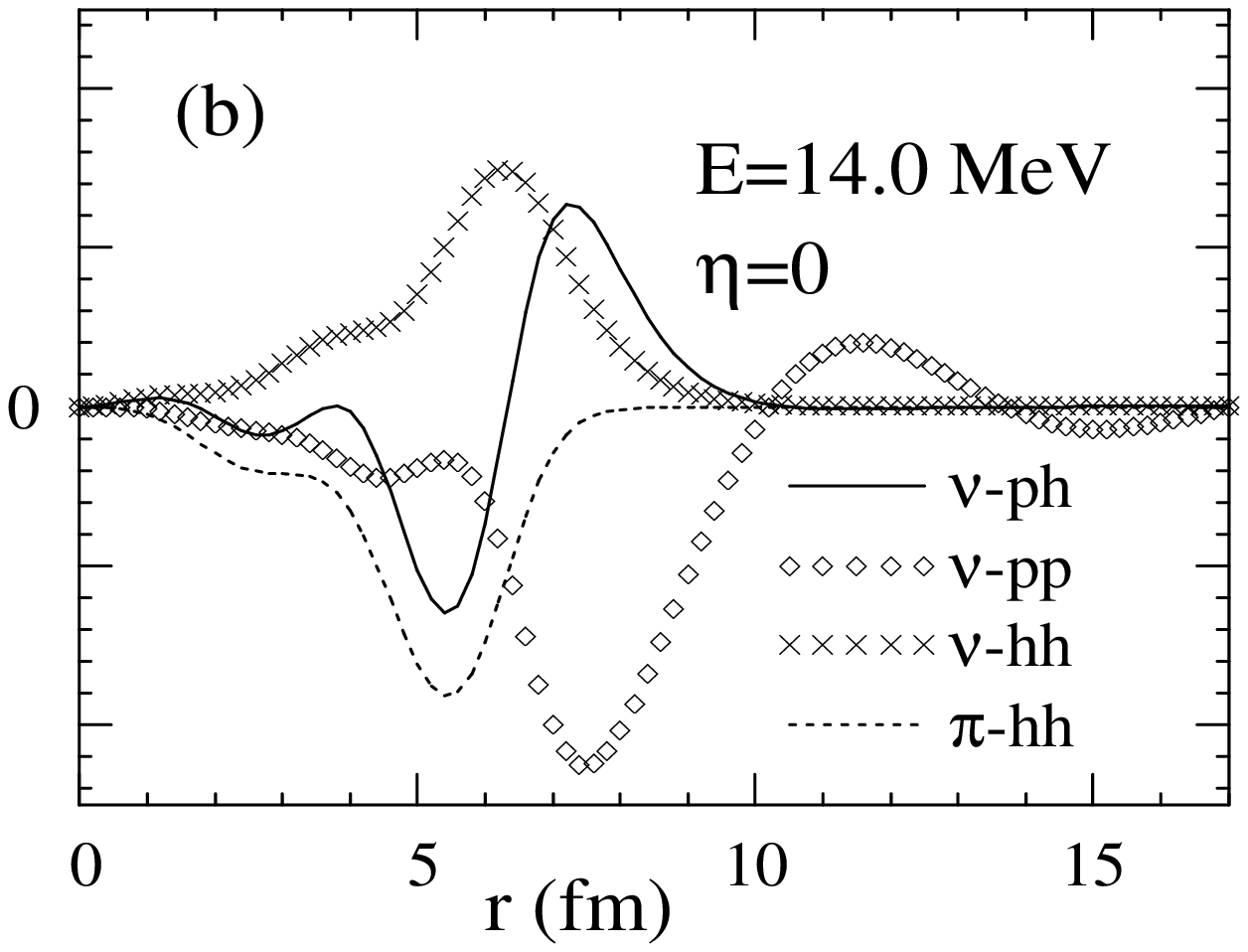}
\end{minipage}
\caption{(a) The same as Fig.5(a), but for the quadrupole response.
(b) The transition densities  evaluated
at the energy $E=14.0$ MeV in the case of strong pairing 
($\eta=0$).}
\label{fig:fig6}
\end{figure}

Fig.6(a) shows the isovector quadrupole response of
$^{158}$Sn for various pairing strengths. There are
three peaks/bump in the strength distribution.
The peak around $E\sim 20-27$ MeV is the
isovector GQR. 
The IV-GQR tends to diminish if
the pairing strength is increased to $\Delta \gesim 7$ MeV.
It is noticed that a low-lying bump around
$E=5-17$ MeV emerges only when the pairing 
is taken stronger ($\Delta > 3$ MeV, $\eta<0.5$). 
(The third peak, the sharp one around $E\sim 2$ MeV
is the surface oscillation 
having an isoscalar character.)
The transition densities of this low-lying bump 
for the strongest pairing $\eta=0$ $\Delta \sim 15$ MeV
is shown in Fig.6(b). The profile is similar to those
of Fig.5(b)(c) in the surface and in the external region, 
and hence this mode is also interpreted as the core-vs-dineutron
mode. Noting the fact that the core-vs-dineutron mode emerges both 
in the dipole and quadrupole responses, we consider
that the core-vs-dineutron mode
is a fundamental mode of excitation in superfluid nuclei
near the neutron drip-line.
In the quadrupole case,  however, 
the low-lying core-vs-dineutron mode emerges only in
the unrealistically strong pairing cases $\Delta \gesim 3$ MeV
while in the dipole case it is present even with the realistic pairing.
This difference is due to the shell effect.

\section{Conclusions}

The spatial di-neutron correlation is closely related to the
pair correlation in the strong coupling regime. 
The neutron pairing in low density uniform matter 
approaches to this situation in a wide density range
$\rho/\rho_0 \sim 10^{-1}-10^{-4}$. The concept of
BCS-BEC crossover is useful to characterize this behavior.

We have discussed how the strong coupling pairing
inluences collective excitations in 
medium-mass neutron-rich nuclei. To obtain an overall picture,
we have varied artificially 
the effective pairing interaction 
so that we can  cover not only a realistic situation, where
the strong coupling pairing is expected only in the surface region
with low density, but also an extreme limit 
where the strong coupling prevails 
in the whole density range and in the whole nuclear volume.

We found that there are two kinds of collective excitation
associated with the strong coupling pairing. The first is
the Anderson-Bogoliubov phonon mode, which is
essentially an hydrodynamic motion of two kinds of superfluid with an 
isovector character. However, this emerges only if an unrealistically strong 
pairing interaction is used. The second is 
the core-vs-dineutron mode taking place around the nuclear surface. 
It shows up in nuclei near the neutron drip-line 
even for the realistic pairing interaction with reasonable 
density-dependence. It is nothing but the soft dipole excitation.  
The soft dipole excitation is an important clue to study
the strong surface pairing in nuclei near the drip-line.

\end{document}